\documentclass[12pt]{iopart}

\usepackage{graphicx}
\usepackage[usenames]{color}

\begin{document}

\title[Atomic Phase Lock]{Accelerating the Averaging Rate of Atomic Ensemble Clock Stability using Atomic Phase Lock}

\author{Nobuyasu Shiga$^{1,2}$, Michiaki Mizuno$^2$, Kohta Kido$^1$, Piyaphat Phoonthong$^{2,3}$, Kunihiro Okada$^4$}

\address{$^1$ PRESTO, Japan Science and Technology (JST), 4-1-8 Honcho Kawaguchi, Saitama 332-0012, Japan}
\address{$^2$ National Institute of Information Communication and Technology, 4-2-1 Nukui-kitamachi, Koganei, Tokyo 184-8795, Japan}
\address{$^3$ National Institute of Metrology Thailand (NIMT),
3/4-5, Moo 3, Klong 5, Klong Luang, Pathumthani, 12120 Thailand}
\address{$^4$ Department of Physics, Sophia University, 7-1 Kioicho,Chiyoda-ku, Tokyo 102-8554, Japan}
\ead{shiga@nict.go.jp}


\begin{abstract}
  We experimentally demonstrated that the stability of an atomic clock improves at fastest rate $\tau^{-1}$ (where $\tau$ is the averaging time) when the phase of a local oscillator is genuinely compared to the continuous phase of many atoms in a single trap (atomic phase lock).  For this demonstration, we developed a simple method that repeatedly monitors the atomic phase while retaining its coherence by observing only a portion of the whole ion cloud.  Using this new method, we measured the continuous phase over 3 measurement cycles, and thereby improved the stability scaling from $\tau^{-1/2}$ to $\tau^{-1}$ during the 3 measurement cycles.
  This simple method provides a path by which atomic clocks can approach a quantum projection noise limit, even when the measurement noise is dominated by the technical noise.

\end{abstract}


\section{Introduction}
\begin{figure}[htbp]
  \begin{center}
    \includegraphics[keepaspectratio=true,width=100mm]{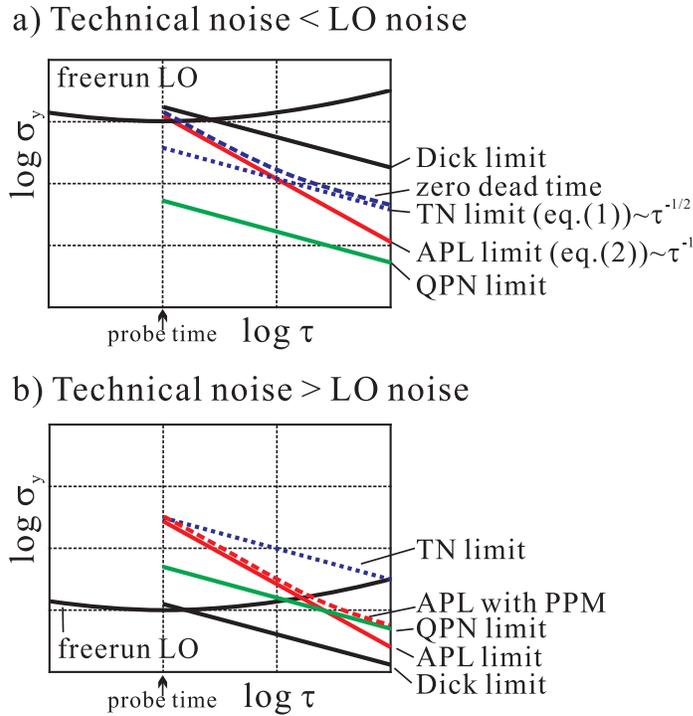}
  \end{center}
  \caption{Typical stability of free-running LO and passive atomic clock.  On both axes, the grid lines indicate decades.  Technical noise(TN) limit and atomic phase lock (APL) limits are given by Eq.(\ref{eq:Sigma1}) and Eq.(\ref{eq:SigmaAPL}), respectively.  Here, we assume TN limit is worse than QPN limit.  a) LO noise exceeds the technical noise.  Free-running LO noise is larger than the TN limit at $\tau=T_c$.  Starting from the free-running noise, $\sigma_y$ decreases at rate $\tau^{-1}$ until it reaches the SNR limit (blue dotted line).  APL follows the same slope but is not limited by the SNR limit.  b) Technical noise exceeds the LO noise.  Since the Dick limit is negligible in this case, the APL alone is responsible for the $\tau^{-1}$ dependence.}
  \label{fig:ClockLimit}
\end{figure}

Much effort has been invested in improving the stability of atomic clocks, which were first demonstrated more than 50 years ago \cite{Essen-Nature-1955}.
The stability of a clock, denoted by $\sigma_y$ (where $y$ is a fractional frequency), is expressed as \cite{Riehle-Book-2004}
\begin{equation}\label{eq:Sigma1}
  \sigma_y=\frac{1}{K\cdot Q\cdot \textrm{SNR}\cdot \sqrt{n}}=\frac{1}{K\cdot Q\cdot \textrm{SNR}}\sqrt{\frac{T_c}{\tau}}
\end{equation}
where $K$ is a constant of the order unity that depends on the spectrum shape, $Q$ is the quality factor, $n$ is the total number of measurements, $T_c$ is the cycle time and $\tau$ denote the total measurement time.
SNR is the Signal-to-noise ratio of a single measurement of an atomic population ratio.
The measurement noise in SNR may comprise in the technical error of detection, quantum projection, fluctuating local oscillator (LO) frequency, fluctuating atomic frequency shifts, and loss of atomic coherence.
The technical noise is a combination of many noises that is related to detection of the signal, including laser power and frequency fluctuation, photon shot noise, atom number fluctuations, and electronics noise.
We note that we mainly discuss about the technical noise, quantum projection noise(QPN) and LO noise, assuming that all other noises are sufficiently smaller in this paper.
The description of clock stability expressed in Eq. \eref{eq:Sigma1} improves the clock stability with $\tau^{-1/2}$ scaling.  When the measurement noise is dominated by technical noise or QPN, the stability line expressed as Eq. \eref{eq:Sigma1} is hereafter referred to as the ``technical noise limit,'' or ``QPN limit,'' respectively.
\Fref{fig:ClockLimit} illustrates the stability of the technical limit and QPN limit, assuming technical limit is larger than QPN limit.

From Eq. \eref{eq:Sigma1}, we observe that stability can be improved by 1) increasing the Q of the resonance by increasing the probing time or carrier frequency, 2) increasing the SNR, and 3) averaging over many measurements (i.e., increasing the $n$).  Past improvements have focused on strategies 1) and 2).  For example, the Q of the optical ion clock developed by Bergquist et al. \cite{Diddams-Science-2001} was several orders of magnitude higher than that of microwave clocks.  Katori et al. \cite{Katori-Nature-2005} demonstrated an optical lattice clock that simultaneously increases the Q and SNR.  Squeezed \cite{Wineland-PRA-1992} or entangled states \cite{Bollinger-PRA-1996} were proposed to improve the technical noise, reducing the technical noise to below the QPN limit\cite{Itano-PRA-1993}.
We need to be careful that Eq. (\ref{eq:Sigma1}) does not account for the effect of the dead time, known as the Dick effect\cite{Dick-PTTI-1987}.  The dead time is the time expended in any process other than probing the atom with microwaves.  The stability is limited by the Dick effect when the frequency noise of the LO is large compared to the technical noise\cite{Itano-PRA-1993} and the dead time is significant.  In this case, reducing the dead time reduces the $\sigma_y$ by $\sigma_y\propto\tau^{-1}$ until the SNR limit line is reached (blue broken line in  \Fref{fig:ClockLimit} (a)).  This idea \cite{Dick-PTTI-1990} has been recently demonstrated \cite{Kasevich-PRL-2013}.

To accelerate the averaging rate of strategy 3), we have proposed an atomic phase lock (APL)\cite{Shiga-NJP-2012}.
Principally, the atomic phase lock lowers the stability at fastest rate $\tau^{-1}$, by genuinely monitoring the phase of the atom.  Provided that the atomic phase remains coherent and is monitored as such, the stability of the atomic clock should improve by $\tau^{-1}$.  However, the stability of atomic clocks normally improves by $\tau^{-1/2}$ even when employing the Ramsey sequence\cite{Ramsey-Book-1956}, which measures the atomic phase.  This trend occurs because the atomic phase is destroyed (and initialized) after each projection measurement cycle.  If the atoms could maintain its phase coherence over many measurement cycles, the stability would reduce much more rapidly (as $\tau^{-1}$), and the stability could be expressed as
\begin{equation}\label{eq:SigmaAPL}
  \sigma_y^{APL}=\frac{1}{K\cdot Q\cdot \textrm{SNR}\cdot n}\sim\frac{1}{K\cdot f_0\cdot \textrm{SNR}\cdot \tau}.
\end{equation}
where $\sigma_y^{APL}$ is the frequency stability under maintenance of the atomic phase coherence.

\Fref{fig:ClockLimit} a) shows the typical stability of an atomic clock that is limited by LO noise.  In the presence of significant dead time, the stability is limited by the Dick limit\cite{Dick-PTTI-1987}.  As already mentioned in the previous paragraph, if the dead time is sufficiently small, then the stability reduces as $\tau^{-1}$ until it reaches the SNR limit.  This means that when limited by LO noise (\Fref{fig:ClockLimit} a)), it is not an ideal situation to demonstrate the APL, because observing the $\tau^{-1}$ could just mean the dead time is small enough.

In order to avoid this ambiguity, we have decided to stay in the technical noise limit, where technical noise is larger than LO noise (\Fref{fig:ClockLimit} b)), for demonstration of APL.  In this case, the observation of the $\tau^{-1}$ dependency only comes from the effect of APL and not from elimination of Dick effect.

We developed a method that projects only a part of atoms, in order to maintain the coherence of an atomic phase for multiple cycles, and we call this method partial projection measurement (PPM).  In principle, the stability ends up the same value whether you perform a projection measurement part by part or all at once.  In other words, projecting $1/n_{cp}$ part of atoms at a time maintains the coherence of phase for up to $n_{cp}$ cycles and reduces the stability at fastest rate by $\tau^{-1}$ for the same $n_{cp}$ cycles, but SNR is reduced by a factor $1/\sqrt{n_{cp}}$ in return.  Therefore, APL using PPM results in the same stability as the conventional method where a projection measurement is performed once for all the atoms.
In this sense, this paper demonstrates a proof-of-principle experiment of APL, but not the actual improvement of the overall clock performance.
We note, however, that APL opens a way to overcome the technical noise limit when an atomic clock's SNR is limited by technical noise.
For example, if we trap one million atoms but we can project only 10000 atoms at one time, the stability is limited by SNR of 100 which is a factor 10 worse than QPN limit for one million atoms.  If we introduce APL using PPM, we maintain the phase lock of LO to atoms for up to 100 cycles of partial projections, and the stability approaches QPN limit for the one million atoms.

The present paper experimentally demonstrates the APL.
Section 2 describes our experimental setup.  We used a single ensemble of ions, and performed projection measurements on only a portion of the ions at a time (Section 3).  Section 4 shows the results of three phase measurements obtained by the proposed method without resetting the atomic states.  The stability is reduced by $\tau^{-1}$ instead of $\tau^{-1/2}$ and the long term stability line is lowered by a factor of $\sqrt{3}$ as expected.  Section 5
discuss the application of the method and suggests ideas for further improvements.

\section{\label{ch:Experiment}Experimental setup}

\subsection{\label{IonTrap}Ion trap}
\begin{figure}[htbp]
  \begin{center}
    \includegraphics[keepaspectratio=true,width=100mm]{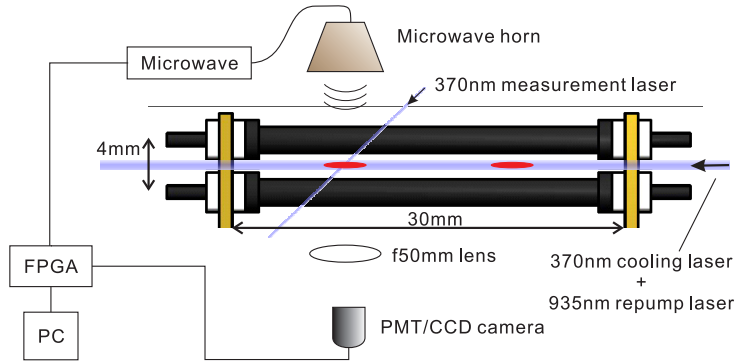}
  \end{center}
  \caption{Schematic diagram of ion trap and measurement system.  Ion clouds are indicated with red oval.}
  \label{fig:IonTrap}
\end{figure}

This section describes our experimental setup.
\Fref{fig:IonTrap} shows our linear radio frequency (rf) quadrupole trap, in which four cylindrical rods (diameter= 2 mm) are placed at the corner of a square separated by 4 mm (center to center) and the DC bias plates are spaced by 30 mm.  Sinusoidal voltage of 10.15 MHz with amplitude 330 V$_{pp}$ and constant 50 V are applied on rf rods and DC bias plates, respectively.  Ytterbium ions of 171 isotope ($^{171}$Yb$^+$) are selectively trapped by a 399 nm photo-ionization laser (not shown in \Fref{fig:IonTrap}) and a 370 nm cooling laser.

Unintentionally, ions were trapped in two locations, delineated by red ovals in \Fref{fig:IonTrap}.
Each cloud (3 mm length) contains about 2000 ions.
Finite element method simulations revealed that our rf rods are much more closely spaced compared to the distance between two bias plates.  At such small separation, the DC potential near the trap center is shielded by the rods.  Consequently, a small potential barrier (whose origin is unclear) remains in the center even with 50 V DC potential on the bias plates.
One of the clouds was used for clock measurements.

\subsection{\label{IonCooling}Ion Cooling}
\begin{figure}[htbp]
  \begin{center}
    \includegraphics[keepaspectratio=true,width=60mm]{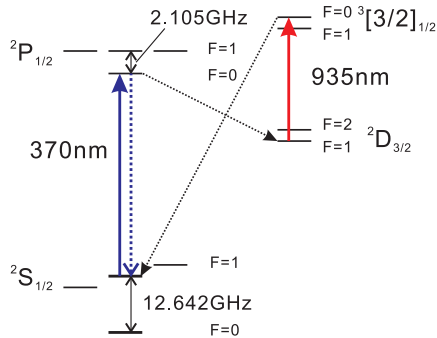}
  \end{center}
  \caption{Simplified energy level diagram of $^{171}\textrm{Yb}^+$.}
  \label{fig:YbIonEnergy}
\end{figure}

Once trapped, the ions are cooled to about 50 mK by Doppler laser cooling.  This 50 mK temperature was obtained by separately measuring the 370nm cooling transition broadening\cite{Phoonthong-2013}.
The ions were predominantly cooled by the 370 nm laser (13 $\mu$W) and the cooling cycle was closed by combining the 935 nm repump laser (6 mW) with 14GHz modulation of the 370 nm cooling laser  \cite{Olmschenk-Thesis-2009} (\Fref{fig:YbIonEnergy}).
The trap area of the vacuum chamber was covered by a single layer magnetic shield of shielding factor 15, yielding an interior field of 0.04 Gauss.  During microwave proving, this residual field was canceled by 3 pairs of Helmholtz coils. The dark states were destabilized\cite{Berkeland-PRA-2002} by a 0.4 Gauss field tilted 45$^\circ$ from the laser axis, using the Helmholtz coils.  The ions were initialized to the $^2\textrm{S}_{1/2}(F=0)$ state by a cooling laser that is phase-modulated at 2.1 GHz by an electro optical modulator.

\subsection{\label{Microwave}Microwave transition}
The 12 GHz microwave clock transition is the hyperfine splitting of the ground state.
The ions were coupled to microwaves emitted by a microwave horn.  The 12GHz microwave synthesizer was referenced to a hydrogen maser, and the synthesizer can be switched between two phase profiles (in our case, between 0$^\circ$ and 90$^\circ$) using external logic.  Microwave emission was terminated by a 60 dB isolation PIN switch.

The population ratio of the ions in the excited state ($^2\textrm{S}_{1/2}(F=1, m_f=0)$) was measured by the electron shelving technique\cite{Dehmelt-PRL-1986}, using the 370 nm laser without 14 GHz modulation.

Timings of switching lasers, microwave amplitude and phase, 14 GHz modulation and 2 GHz modulation of the cooling laser were precisely controlled by field programable gated array (FPGA).  The FPGA also counted the number of photons, and passed the data to PC for further data processing.

\section{\label{Ch:APLViaPPM}APL via partial projection measurement}
\subsection{\label{PPM}Partial projection measurement}
We now introduce partial projection measurement (PPM).
In PPM, a 16 $\mu$W diagonal laser is operated at 370 nm (see \Fref{fig:IonTrap}) such that only a portion of the ions interacts with the laser.  The waists of the ions and diagonal beam are about equal (approximately 200 $\mu$m).  The basic measurement sequence is as follows.
\begin{enumerate}
  \item Initialize atomic state (cooling and pumping to $^2\textrm{S}_{1/2}(F=0)$).
  \item Manipulate state with microwaves (details are shown in Figures \ref{fig:PPRabi} and \ref{fig:PPM})
  \item Perform PPM
  \item repeat steps (ii)-(iii).
\end{enumerate}
After the first partial measurement, the projected ions became mixed with un-projected ions during manipulation period prior to the next measurement.  We note that our ion temperature was sufficiently high enough such that the ion cloud never crystalized.  The measured population is valid provided that the ratio of un-projected to projected ions exceeds the SNR of the measurement.

For example, when there are 10,000 total ions and 100 ions are measured at SNR=10 via PPM, 1 \% of the total ions are projected at each PPM and will give wrong information of the phase in the next measurement.  Since this projected ions scatters in the whole ion cloud during step (ii), the ratio of projected ions increases exponentially as $1-0.99^{n_{cp}}$, as one repeats the PPM for $n_{cp}$ times.  Since this projected ions results in phase estimation error, one should limit the $n_{cp}$ to less than 10 measurements, so that the error due to projection is less than the technical noise.  In other words, we maintain SNR constant by terminating the APL before the noise due to the projection exceeds the technical noise.

\subsection{Check of PPM with Rabi oscillation signal}
\begin{figure}[htbp]
  \begin{center}
    \includegraphics[keepaspectratio=true,width=80mm]{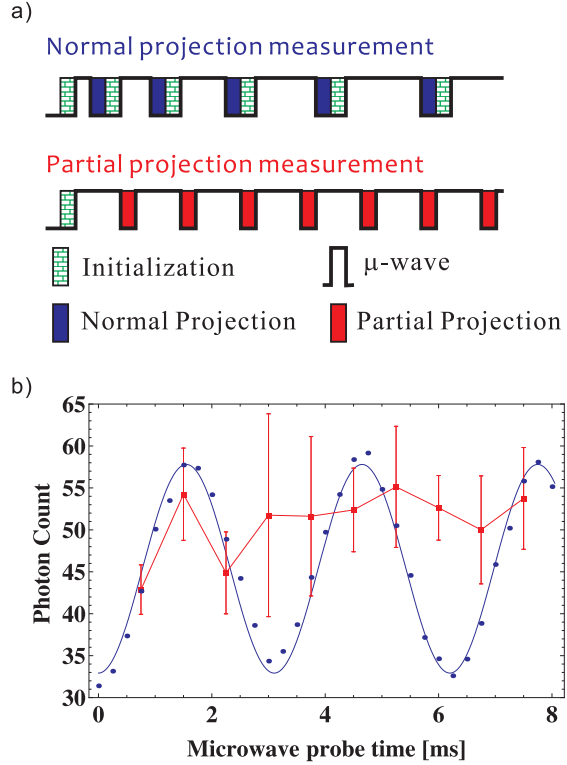}
  \end{center}
  \caption{a) Sequence of normal projection measurement and PPM. b) Cases in which the ion state is re-initialized and not re-initialized after each measurement are shown in blue and red, respectively.  Filled blue circles are the averages of 10 measurements.  Solid line is a sinusoidal fitting.  Filled red squares are the averages over 8 measurements.  The error bars represent the standard deviations in a single measurement.}
  \label{fig:PPRabi}
\end{figure}

The validity of PPM was tested by first measuring the Rabi oscillation under PPM.  This test evaluates whether PPM can correctly measure a population that is constantly changed by microwave interaction.  \Fref{fig:PPRabi} a) shows the measurement sequence of normal Rabi oscillations obtained under PPM.
In standard Rabi oscillation measurement, the atomic state is reset after each measurement cycle, and the probe time is increased at each cycle.  In the PPM approach, the atomic state is not reset, rather another rotation is added in each cycle.
The measured result is shown in \Fref{fig:PPRabi} b).  If PPM is valid, the red data should align with the blue sinusoidal curve.  We observe that PPM is correct up to three measurements, but desynchronizes from the correct population at and beyond the 4th measurement.  This decoherence is further discussed in the next section.

\subsection{Check of PPM with continuous phase measurement (modified Ramsey method)}
We now test the decoherence due to PPM in phase measurement sequences.
To monitor the total phase difference accumulated over multiple measurement cycles, we must modify the Ramsey sequence as well as the projection measurement.  Our modified Ramsey sequence is shown in \Fref{fig:PPM} a) and proceeds via the following steps:
\begin{figure}[htbp]
  \begin{center}
    \includegraphics[keepaspectratio=true,width=80mm]{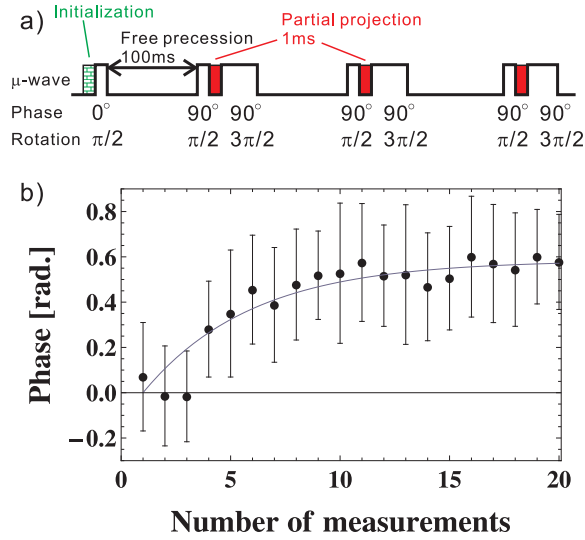}
  \end{center}
  \caption{a) Sequence of continuous phase measurement with PPM. b) Continuous phase measurement (up to 20 PPM) with diagonal laser beam.  Filled circles are averaged over 32 samples and error bar indicate the standard deviation.}
  \label{fig:PPM}
\end{figure}
Basic sequence is,
\begin{enumerate}
  \item Initialize once at the beginning of the measurement,
  \item Construct a superposition state by applying a $\pi/2$ rotation (0.75 ms) with 0$^\circ$ phase,
  \item Wait for a free precession time,
  \item Transfer the phase to the population by applying a $\pi/2$ rotation with 90$^\circ$ phase,
  \item Perform PPM,
  \item Revert the population to the original phase by applying a $3\pi/2$ rotation with 90$^\circ$ phase.
  \item repeat (iii) $\sim$ (vi).
\end{enumerate}
This sequence is very similar to that proposed in our previous paper \cite{Shiga-NJP-2012}, but is slightly modified to accommodate technical limitation.  Our 12GHz signal generator can switch phase with the external logic control in only two profiles (in our case, at 0$^\circ$ and 90$^\circ$),  so the 1/2$\pi$ rotation at the 270$^\circ$ phase was replaced with a 3/2$\pi$ rotation at the 90$^\circ$.  The 12 GHz microwave signal generator was referenced to a hydrogen maser, and the frequency noise was much smaller than the SNR of the measurements.  The phase error in a hydrogen maser at 0.1 s averaging time is below 0.02 radians.  This implies that the measured zero phase shift between the LO and atoms should always lie within the measurement noise.  \Fref{fig:PPM} b) shows the measured phase over 20 PPM measurements.  Each datum is the average of 32 measurements and measurement noise ($\sim$0.2 radians) is mainly due to scattering of the 370nm laser.
Clearly, PPM measures the correct phase over three consecutive measurements, and thereafter deviates from the true phase by more than one sigma.  After the 3rd measurement,the atomic phase is abruptly loses coherence and the population ratio rapidly asymptotes toward 70\% excitation.  Together with the Rabi oscillations measured by PPM, we conclude that our continuous phase measurement is valid up to three measurements.

We consider a simple model, in which a constant number of ions are projected and lose coherence during each measurement.  Prior to the $n$-th measurement, the proportion of ions whose states are projected, denoted as $P_{proj}$, is given by,
\begin{equation}\label{eq:decoherence}
  P_{proj}=1-(1-p)^{n-1}
\end{equation}
where $p$ is the ratio of the number of ions that get projected in a single measurement, normalized by the total number of ions.  Fitting the data in \Fref{fig:PPM} to Eq. (\ref{eq:decoherence}) and an additional fitting parameter (the amplitude) yields the solid curve in \Fref{fig:PPM}.  From this fitting, $p$ is estimated as 18 \%.

\subsection{Simulation evaluation of PPM}
To elucidate the decoherence rate, we simulated the ion motions using a molecular dynamics method\cite{Okada-PRA-2010}.  We calculated the motions of 2000 ions trapped in a potential that corresponding to our trap parameters, assuming constant temperature.  We confirmed that ions undergo Brownian motion with a mean-squared displacement along the optical axis $\Delta z^2\equiv \langle(z-z_0)^2\rangle$ (where $z_0$ is the initial position at $t=0$), given by
\begin{equation}
  \Delta z^2=2D t
\end{equation}
where $D$ is the diffusion constant of ions and $t$ denotes time.

Simulating this ensemble with T=50 mK, we obtained $D=3.5\times 10^{-6} [m^2 s^{-1}]$.
Varying the temperature from 10 mK to 2 K, we found that D and T were related through the Boltzmann constant and the mobility $\mu$, namely,
\begin{equation}
  D=\mu k_B T.
\end{equation}
The fitting gives $\mu=8.62\times10^{18} [m^2s^{-1}J^{-1}]$.

Next, we counted the number of ions passing through the region in which the diagonal laser and ions overlap.  At $T= 50$ mK, 17 \% of the total ions were struck by the measurement laser within 1 ms.  This implies that 17 \% of the ions were projected and decohered during 1 ms measurement time, consistent with the estimates of \Fref{fig:PPM}.  The cause of the abrupt decoherence after the 3rd measurement remains unclear.

\section{\label{ch:Allan}Allan deviation of APL based atomic clock}
\begin{figure}[htbp]
  \begin{center}
    \includegraphics[keepaspectratio=true,width=80mm]{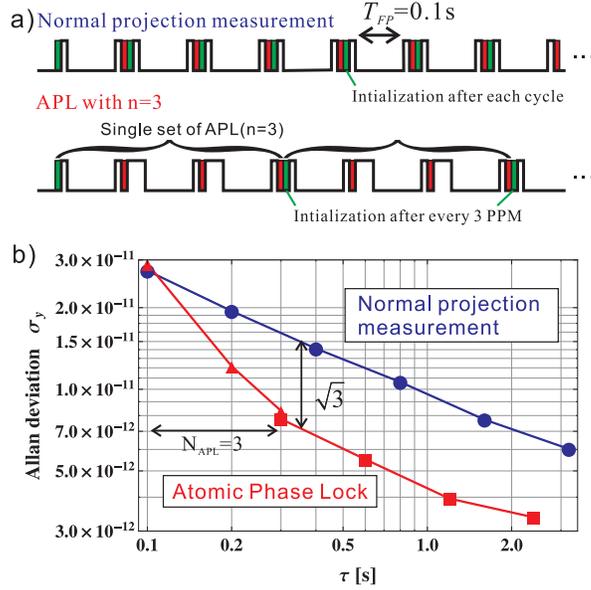}
  \end{center}
  \caption{Standard deviation of n-th APL measurements (red filled triangles) and the Allan deviation (red filled squares) in the 3rd measurement of APL cycles.  Allan deviations of the regular projection measurements (blue filled circles) are also shown.}
  \label{fig:APLAllan}
\end{figure}

This section experimentally compares the stability of the standard method (in which phase is initialized during each cycle) with that of APL.  APL initializes the phase after each sequence of 3 PPMs, as shown in \Fref{fig:APLAllan} a).
\Fref{fig:APLAllan} b) shows the standard deviation in the APL results (filled red triangles) and the overall Allan deviation (filled red aquares).
APL ($\tau=0.1\sim0.3$) and APL repetition ($\tau>0.3$) are calculated in different ways.
For $\tau=0.1\sim0.3$, frequency error $\Delta f$ of LO compared to atomic frequency is estimated from the measured total phase $\phi_n$ as,
\begin{equation}\label{DeltaF}
  \Delta f = \frac{\phi_n}{n T_{FP}}
\end{equation}
where n indicates the n-th APL measurement cycle ($n=$1, 2, or 3) and $T_{FP}=0.1 s$ is a free precession time of the Ramsey sequence.  In \Fref{fig:APLAllan} b), the standard deviation in the n-th measurement scales as $\tau^{-1}$ (red triangles).
For $\tau>0.3$, APL cycles are established, and we calculate and plot the Allan deviation from the final (3rd) measurement in each cycle (red squares).
The triangles and circles in \Fref{fig:APLAllan} b) are calculated from the same data and the slight mismatch at $\tau=0.3 s$ is due to the difference between the standard and Allan deviations.

For comparison, the Allan deviation of the regular Ramsey measurement, in which the phase is initialized during each cycle, is also plotted in \Fref{fig:APLAllan} b).  This deviation scales as $\tau^{-1/2}$.  Based on the measured Allan deviations, the stability of APL is improved by a factor of $\sqrt{3}$, relative to the standard Ramsey cycles. Overall, we have demonstrated that APL improves the stability by a factor of $\sqrt{n_{cp}}$, where $n_{cp}$ is the number of continuous APL measurement.

The number of valid PPMs can be increased by 1) reducing the temperature of the ions (lowering the Diffusion coefficient), 2) trapping more ions (reducing the proportion of the decohered ions in a single measurement), or 3) adopting weak measurements.  Since weak measurement ideally allows us to estimate phase but yet to preserve the phase over multiple measurement cycles, we expect that this last strategy can greatly increase the number of valid PPM.  A promising weak measurement scheme is Faraday Rotation,discussed in detail in our previous paper\cite{Shiga-NJP-2012}.

\section{\label{Discussion}Discussion}
%
Although this experiment isn't a demonstration of the actual improvement of a clock performance, as already mentioned, 
APL via PPM will be useful when the performance of an atomic clock is limited by technical noise.
In the past, trapping atoms more than (SNR)$^2$ was futile.  APL via PPM opens a path to utilize more number of atoms than (SNR)$^2$ for better stability, thereby lowering the stability line in \Fref{fig:ClockLimit} to
\begin{equation}\label{eq:AllanAPL}
  \sigma_y=\frac{1}{\sqrt{n_{cp}} K\cdot Q\cdot \textrm{SNR}}\sqrt{\frac{T_c}{\tau}}.
\end{equation}
PPM limits the $n_{cp}$ to $N_{atom}/(\textrm{SNR})^2$ (where $N_{atom}$ is the total number of atoms in the trap) because the number of measured atoms is adjusted to (SNR)$^2$ and all of these atoms become decohered during a single measurement.  When $n_{cp}=N_{atom}/(\textrm{SNR})^2$, Eq. \eref{eq:AllanAPL} computes the QPN limit line.  Therefore, when APL is based on PPM, the system cannot be improved beyond the  QPN limit (\Fref{fig:ClockLimit}),
except perhaps by weak measurement.  However, discussion of the weak measurement limit is beyond the scope of this paper.

Since we performed the proof-of-principle experiment in the regime that is limited by technical noise, we didn't have to feed the atomic signal back to the LO.  Our next step would be to perform the APL for the case where LO noise limits the stability, in order to demonstrate the actual improvement of a clock performance.  In that case, we would need to evaluate the servo error in the feed back system to the LO frequency carefully.

Recently, use of multiple atomic traps has been proposed \cite{Sorensen-PRL-2013-2, Rosenband-arXiv-2013}.  This scheme shows the same $\tau^{-1}$ scaling and an overall stability is reduced by a factor of $m2^{-m}$, where $m$ is the number of atomic traps.  Further stability improvements are conceivable if this scheme could be combined with APL.

\ack
This work was supported by the JST PRESTO program and NICT.  We thank H. Hachisu and T. Ido for comments on this manuscript.

\section*{References}
\bibliographystyle{unsrt}

\end{document}